\documentclass[pra,aps,showpacs,showkeys,groupedaddress,superscriptaddress,twocolumn,longbibliography]{revtex4-2}

\usepackage{amsmath}
\usepackage{amssymb}  
\usepackage{bbold}    
\usepackage{braket}
\usepackage{graphicx}
\usepackage{subcaption}
\usepackage{fix-cm}
\usepackage{pgfplots} 
\pgfplotsset{compat=1.5} 
\usepackage{color}
\usepackage{float}
\usepackage{ulem}
\usepackage[colorlinks=true,linkcolor=blue,citecolor=blue,urlcolor=blue]{hyperref}  
\usepackage{todonotes}

\usepackage{array}

\usepackage{bm}	
\renewcommand{\vec}[1]{\bm{#1}}

\begin{document}
\title{An Autonomous Topological Pump}

\author{Julius Bohm}
\affiliation{Department of Physics and Research Center OPTIMAS, RPTU, University Kaiserslautern-Landau, 67663 Kaiserslautern, Germany}
\author{James Anglin}
\affiliation{Department of Physics and Research Center OPTIMAS, RPTU, University Kaiserslautern-Landau, 67663 Kaiserslautern, Germany}
\author{Michael Fleischhauer}
\affiliation{Department of Physics and Research Center OPTIMAS, RPTU, University Kaiserslautern-Landau, 67663 Kaiserslautern, Germany}
\affiliation{Research Center QC-AI, RPTU,  University  Kaiserslautern-Landau, 67663 Kaiserslautern, Germany}


\begin{abstract}
Robust quantization of particle transport, as in a Thouless pump, is a hallmark of topological quantum systems with externally controlled system parameters. Here we instead propose and analyze a Thouless pump, for fermions in a one-dimensional lattice, in which external control is not needed, because an additional dynamical degree of freedom allows the pump to work autonomously. The external control parameters are replaced by a quantum spin in a static magnetic field, so that Larmor precession of the spin performs the control cycle that induces topologically quantized transport of the fermions---at least in some higher energy eigenstates of the combined system. In other states, the back-action of the fermions on the spin can distort the control cycle enough to disrupt the transport, but we find numerical evidence for a critical value of the magnetic field above which the autonomous pump works with topological robustness, suggesting that topological protection and autonomous operation together may permit robust "quantum motors".
\end{abstract}

\date{\today}
\maketitle


\section{Introduction}

The realization of microscopic quantum engines 
\cite{kosloff2014quantum,cangemi2024quantum} that are autonomous, rather than being driven by control parameters with externally fixed time dependence, is a challenging goal of quantum technology and an important subject in the growing field of quantum active matter \cite{adachi2022activity,yamagishi2024proposal,takasan2024activity,khasseh2025active,nadolny2025nonreciprocal,penner2025heat,antonov2025engineering}. Whereas any engine needs to sustain a non-equilibrium dynamical process converting energy resources into useful work, an autonomous engine must also achieve this without external control, by regulating its own dynamics.
Achieving autonomous engine operation in the microscopic world, where thermal and quantum fluctuations are large, is an outstanding challenge. Pumps that do not necessarily perform work, but that do induce transport, are a first step towards quantum engines; robust particle transport, in spite of imperfect control, can be attained by exploiting topology. Here we extend the concept of topological pumping to autonomous systems, by presenting a Thouless pump \cite{thouless_pump} in which the externally time-dependent control parameters are replaced by dynamical degrees of freedom, with the Hamiltonian of the entire system being time-independent.

A Thouless pump exhibits quantized particle transport in an insulating bulk state of a 1D lattice upon cyclic adiabatic changes of system parameters.
The transport is governed by an integer topological invariant, equivalent to a Chern number, and is therefore robust against disorder and fluctuations \cite{Niu_1984}. 
Experimentally, these systems can be realized in cold-atom experiments \cite{lohse2016thouless,nakajima2016topological,lu2016bec,minguzzi2022floquet} as well as in photonic systems \cite{kraus2012pumping, verbin2015pumping, jurgensen2021quantized, jurgensen2023quantized}. 
In all previously proposed and realized experimental platforms, however, Thouless pumps have always needed an explicit time-dependent modulation of the fermion system parameters: the pumps are externally driven rather than being autonomous. Progress toward autonomous operation has been registered in cavity QED experiments \cite{rayleigh1883xxxiii,van1927vii,jenkins2013self,ben2021quantum} in which an open system with constant external drive relaxes spontaneously into a stationary limit cycle that
\begin{figure}[H]
    \centering
    \includegraphics[width=0.85\linewidth]{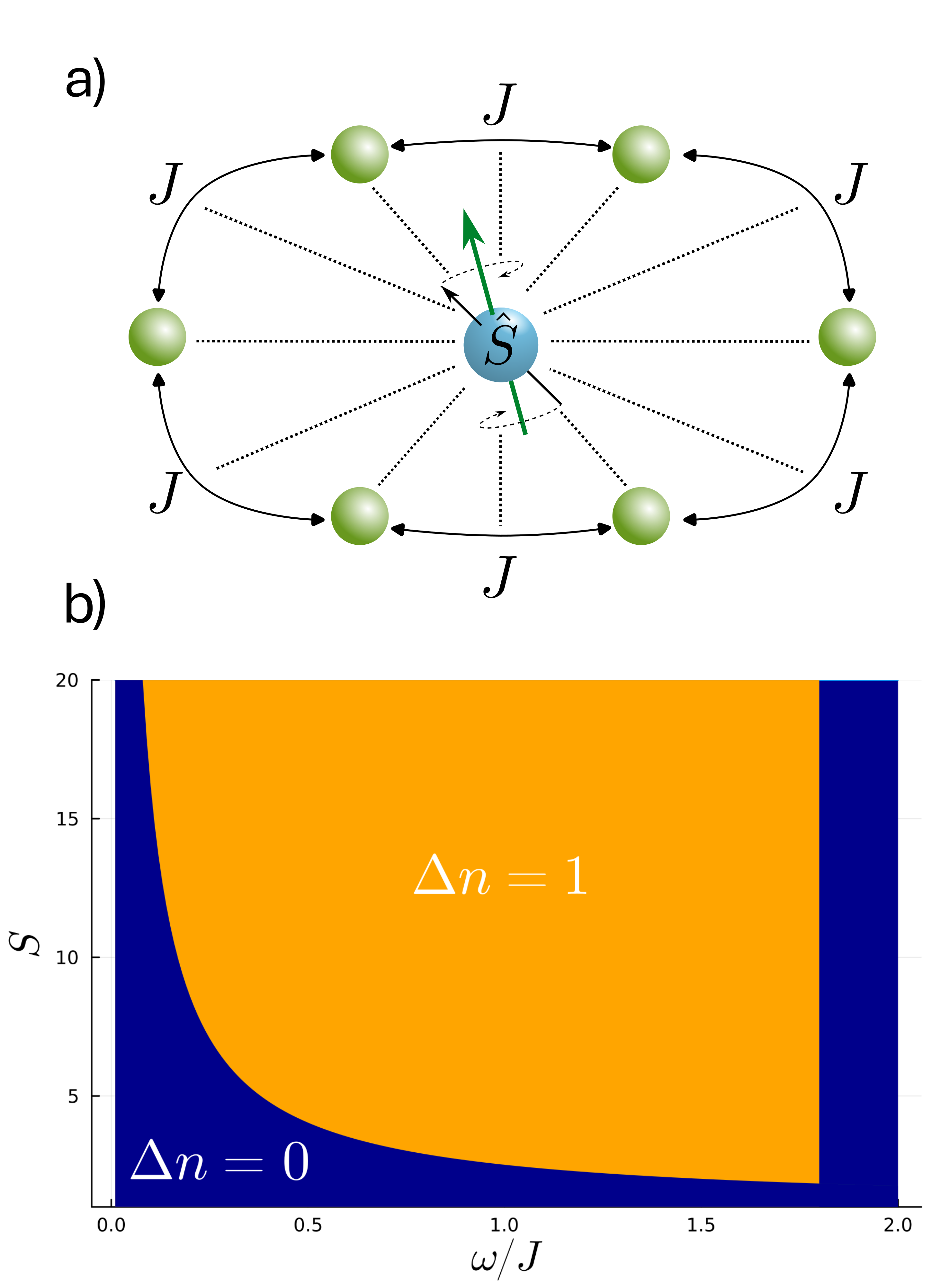}
    \caption{(a) 1D fermion lattice controlled globally by a single (large) spin in a magnetic field $B_z=\omega$. The $\hat S_x$-component controls the hopping amplitude while the $\hat S_y$-component determines the on-site potentials. (b) Schematic phase diagram, showing regions of quantized transport per spin precession time in a specific (excited) eigenstate.}
    \label{fig:system}
\end{figure}
\noindent drives particle transport \cite{dreonSelfOscillating}.

In this paper we theoretically investigate a simple model for a fully autonomous Thouless pump, which operates as a time-independent, isolated system, by incorporating as part of the system an additional quantum degree of freedom whose dynamical time evolution can drive the pump. 
Specifically we consider a generalization of the 
the Rice-Mele-model (RMM) \cite{rice1982elementary} (a well-known example of a Thouless pump in a one-dimensional lattice of fermions), where the time-dependent external parameters are replaced by different components of a central quantum spin in a fixed magnetic field; see Fig.~\ref{fig:system}a). While the RMM or other similar topological pumps are essentially single-particle models and as such fully understood, the coupling to additional dynamical quantum degrees, as necessary for autonomous operation, brings the model into the range of interacting topological systems, which are less well understood.
Our system can therefore serve as a prototypical toy model for autonomous topological pumps in general.

We here show that there are excited eigenstates of our total, time-independent Hamiltonian, which in certain parameter regimes display a steady current of fermions. This current is moreover quantized as an integer multiple of a frequency that is determined by the strength of the external magnetic field, and the quantization is shown to be of the same topological character as the transport in the externally driven Thouless pump. The current in this transporting phase is robust against disorder and perturbations, protected by topology, even though our total Hamiltonian is gapless. 

Whether or not this topological effect actually occurs in any given energy eigenstate of the total system depends on the details of how the whole system interacts. For increasing strength of the magnetic field we first observe a sharp transition of the excited state from a non-transporting state to a transporting state, and then for an even larger field strength we see another transition to a phase without quantized transport; see Fig.~\ref{fig:system}b).  

Our paper is organized as follows. We first briefly explain the RMM, on which our autonomous model is based, and next show how we replace the time-dependent parameters of the RMM with the dynamical quantum degrees of freedom of a large spin (a rotor with a large total angular momentum eigenvalue) in a magnetic field. We then show, by numerical simulations of small systems, that the coupled spin-fermion model possesses excited eigenstates which, at least within a finite range of magnetic field strengths, exhibit
stationary quantized transport of fermions. We give an intuitive explanation of the transport phase diagram, employing a self-consistent mean-field approach, and derive a topological invariant for an approximation to our many-body model which is valid deep in the transporting phase. Finally, we demonstrate the robustness of our transport against disorder.

\section{The model} 

\subsection{Rice-Mele Model}

A paradigmatic example of a Thouless pump is the Rice-Mele-model \cite{rice1982elementary} for fermions with creation and \mbox{annihilation} operators $\hat c_j^\dagger$ and $\hat c_j$, on a one-dimensional lattice, where the index $j$ denotes the lattice site:
\begin{align}
    H_{\mathrm{RM}} = &- \frac{J}{2} \sum_{j=1}^L \Bigl( 1 + (-1)^j \gamma (t)\Bigr)(\hat{c}_j^\dagger \hat{c}_{j+1} + h.c.) \\
    &+ \Delta (t) \sum_{j=1}^L (-1)^j \hat{n}_j. \nonumber
    \label{eq:RMhamiltonian}
\end{align}
The model's key features are a staggered hopping term of strength $J/2(1\pm \gamma(t))$ and an alternating on-site potential $\pm \Delta(t)$, resulting in a two-site unit cell. The lattice therefore in general possesses two energy bands, with the instantaneous energy gap
\begin{align}
    \Delta E_{Gap} = 2 \sqrt{J^2\gamma(t) ^2 + \Delta(t) ^2}.
\end{align}
As long as at least one of the two parameters $J\gamma(t)$ or $\Delta(t)$ is non-zero, then, the system is insulating in the many-body ground state for particle number $N=L/2$, where the whole lower band is occupied and so all excitations require at least $\Delta E_{Gap}$.

A less obvious feature of this half-filled insulating ground state is the effect of adiabatic, cyclic modulation of $\gamma(t)$ and $\Delta(t)$. If the parameters are modulated in any way such that the origin in the 2D parameter space $\{J\gamma(t),\Delta(t)\}$ is encircled,  slowly compared with the band gap frequency, then there is quantized transport of particles in every such cycle.
In an insulating state, it may be surprising that there is any transport at all; transport occurs because the system is time-dependent. Over the duration of a full parameter cycle, the first post-adiabatic correction to time evolution has a finite effect, no matter how slowly the Hamiltonian parameters change.

Concretely, under $\hat{H}_{\mathrm{RM}}$ the current from site $l$ to $l+1$ (with $l=N+1$ being identified with $l=1$ around the ring) is 
\begin{align}\label{RMMJ}
    \mathcal{J}_{l} =  -i\frac{J}{2}\Bigl\langle (1 + (-1)^{l}\gamma(t)\bigl( \hat{c}_{l+1}^{\dagger}\hat{c}_{l} - \hat{c}_{l}^{\dagger}\hat{c}_{l+1}\bigr) \Bigr\rangle\;.
\end{align}
The number of particles transported through the interface between sites $l$ and $l+1$ in a time interval $(t,t+\tau)$ is then given by 
$\Delta n(t,t+\tau) = \int_t^{t+\tau}\!\! d t^\prime\, {\cal J}_l(t^\prime)$.
In particular, with the lower band filled ($N=L/2$) and parameter time dependence such that $(\gamma,\delta)$ winds around the origin $(0,0)$ in every period $T$, as for example with
\begin{align}
    \gamma(t) &= \delta \cos(2\pi t/T), \qquad   \Delta(t) = \delta \sin (2\pi t/T)\;,\label{eq:RM-parameter}
\end{align}
then as long as the cycling frequency $2\pi/T$ is always small enough compared to the instantaneous band gap to maintain adiabaticity, the total particle transport in every period $T$ equals an integer topological invariant:
\begin{align}
    \Delta n= \frac{1}{2\pi} \int_0^T\!\!  dt \, \frac{\partial}{\partial t} \phi_\mathrm{Zak}(t) \quad \in\,  \mathbb{N}.
\end{align}
Here, the so-called Zak phase $\phi_\mathrm{Zak}(t)$ is 
\begin{align}
    \phi_\mathrm{Zak}(t) = \int \mathrm d k  \bra{u_0(k, t)} (-i\nabla_k)  \ket{u_0(k, t)}\;,
\end{align}
while $\vert u_0(k,t)\rangle$ is the instantaneous Bloch state of the band $n=0$ at lattice momentum $k$.

\subsection{Rice-Mele model coupled to a central spin}
\label{sect:mean-field}

Let us now consider a system in which the parameters $\gamma(t)$ and $\Delta(t)$ are replaced by components $\hat S_x$ and $\hat S_y$ of the dimensionless spin operator $\hat{\vec{S}}$:
\begin{align}
    \hat{H} = \hat{H}_{\mathrm{spin}} &- \frac{J}{2} \sum_{j=1}^L \Bigl( 1 + (-1)^j g \hat{S}_x\Bigr) (\hat{c}_j^\dagger \hat{c}_{j+1} + h.c.) \label{eq:mbhamiltonian}\\
    &+ \Delta \sum_{j=1}^L (-1)^j g \hat{S}_y \hat{n}_j. \nonumber
\end{align}
If we associate the $x,y,z$ axes with indices $1,2,3$, the spin components satisfy the angular momentum relations
\begin{align}
    [\hat{S}_i,\hat{S}_j]=i\sum_{k=1}^3\varepsilon_{ijk}\hat{S}_k,\qquad 
    \sum_{k=1}^3 \hat{S}_k^2 = S(S+1)
\end{align}
for a constant (half-)integer $S$. By letting $g =\eta  S^{-1}$ for $0< \eta <1$ we let $\eta$ control the effective size of the control parameter operators for any value of $S$. 
Throughout this paper we will set 
%
$J= \Delta $
%
for simplicity, but generalization to $J\not=\Delta$ is straightforward. A sketch of our system is shown in Fig.~\ref{fig:system}a).

Our total Hamiltonian $\hat{H}$ is time-independent. The time dependence of the control parameters which should drive our pump through its cycle is therefore provided by specifying $\hat{H}_{\mathrm{spin}}$ as that of a magnetic moment in a constant magnetic field, providing the Larmor frequency $\omega$ which should be analogous to the RMM pumping rate $2\pi/T$:
\begin{align}
    \hat{H}_{\mathrm{spin}}= - \omega \hat{S}_z.
\end{align}
For $\omega=0$ our Hamiltonian would have time reversal symmetry, and could thus 
not show a non-vanishing current (unless time reversal were spontaneously broken).
With $\omega\not=0$, on the other hand, the evolution generated in the spin by $\hat{H}_{\mathrm{spin}}$ alone would provide the precession
\begin{align}
    \langle\hat{S_x}\rangle_{t} &= \langle\hat{S_x}\rangle_{t=0}\,  \mathrm{cos}(\omega t) + \langle\hat{S_y}\rangle_{t=0}\,  \mathrm{sin}(\omega t) \label{eq:Sxy_expec_time}, \\
    \langle\hat{S_y}\rangle_{t} &= \langle\hat{S_y}\rangle_{t=0} \, \mathrm{cos}(\omega t) -\langle\hat{S_x}\rangle_{t=0} \, \mathrm{sin}(\omega t), \nonumber
\end{align}
so that at least in the classical correspondence limit $S\to\infty$ we can expect the dynamical evolution of $\hat{S}_{x,y}$ to substitute effectively for the externally time-dependent parameters $\gamma(t)$ and $\Delta(t)$ of the RMM. 

In fact the evolution of $\hat{S}_{x,y}$ is not just driven by $\hat{H}_{\mathrm{spin}}$, but by the full $\hat{H}$ of (\ref{eq:mbhamiltonian}), which includes back-action from the fermions onto the spin. Hence it is not actually guaranteed in general that $\hat{S}_{x,y}$ will simply rotate steadily---and neither are the operators $\hat{S}_{x,y}$ simply equal to their c-number expectation values. Because the transport driven by the Thouless pump is topologically protected, however, we will show that it can persist in spite of back-action and quantum fluctuations.

\section{Eigenstates with quantized fermion current}
\begin{figure}[H]
    \begin{centering}
        \includegraphics[width=0.9\linewidth]{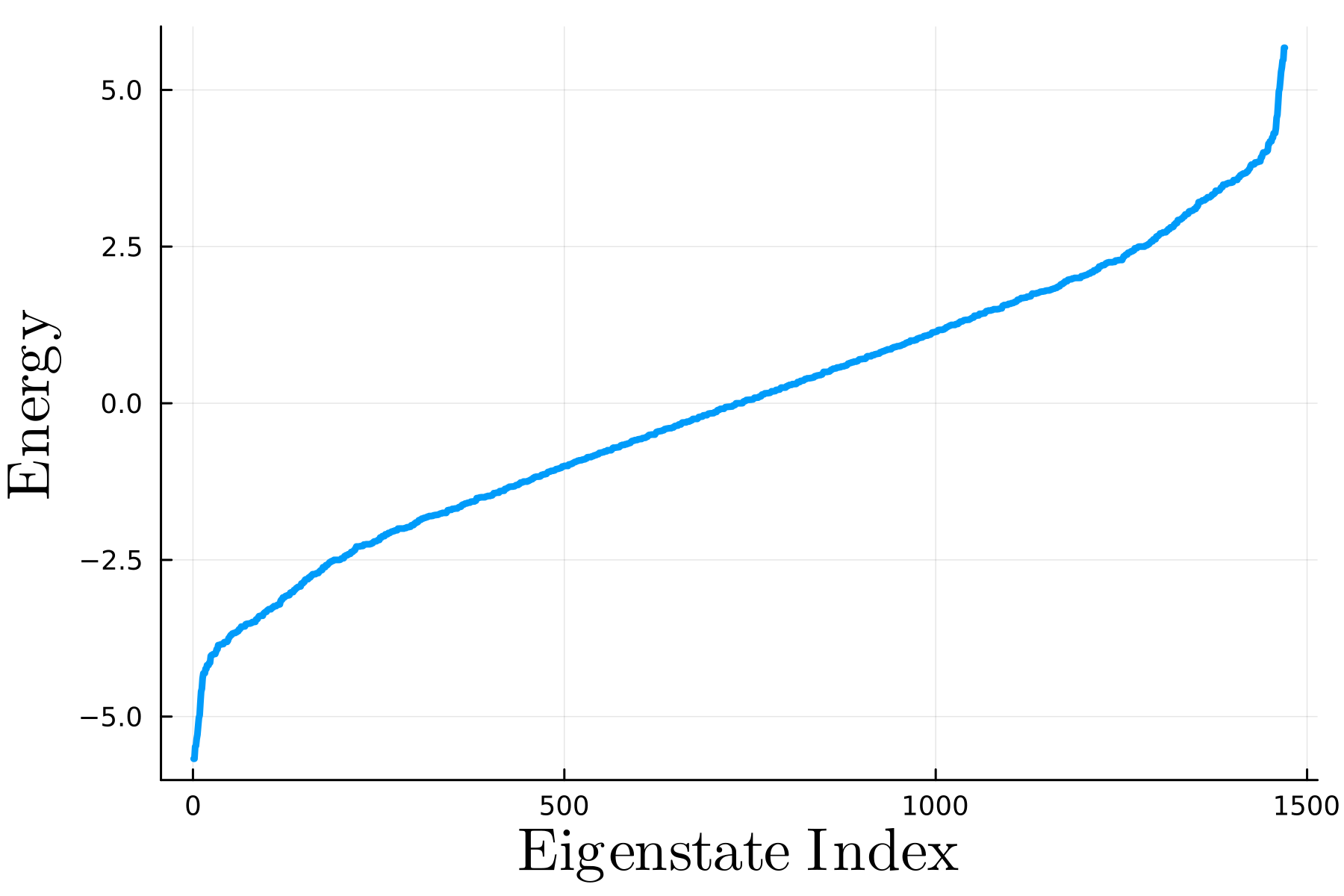}
    \end{centering}
    \caption{Energy spectrum of the full Hamiltonian~\eqref{eq:mbhamiltonian} in units of $J$, in the case of $N=4$ fermions able to hop at rate $J$ between $L=8$ sites, with a spin of size $S=10$, $\eta = 0.95$ and a Larmor frequency $\omega = 0.25 J$. The spectrum is discrete, but shows no gaps when plotted with visibly large dots.}
    \label{fig:fullspectrum}
\end{figure}

Since our total Hamiltonian \eqref{eq:mbhamiltonian} is time-independent, everything which can happen in the model must be determined by diagonalizing $\hat{H}$. For small enough $L$ and $S$ we can do this numerically with exact diagonalization, but identifying the behavior of a topological pump within $\hat{H}$ presents some conceptual challenges, which we will attempt to resolve in a heuristic way. The first such challenge appears immediately: although topological pumping in the RMM is usually understood to depend on the existence of an energetic band gap, the addition of the spin gives our total $\hat{H}$ a gapless spectrum, as shown in Fig.~\ref{fig:fullspectrum}. The finite system does have discrete energies, but the level spacings are too small to be seen in the Figure, and there is no larger gap present. 

On further reflection we may consider that it is only the energies of the fermions which must possess a gap, and not necessarily the whole system. Since $\hat{H}_{\mathrm{spin}}$ does not commute with the rest of $\hat{H}$, ``the energy of the fermions'' is not a good quantum number; eigenstates of $\hat{H}$ are not fermion energy eigenstates. 
So how can we identify which states are the topological pumping states?
For each eigenstate $|\psi_n\rangle$ of $\hat{H}$ we can compute the expectation value of the fermion energy $\hat{H}-\hat{H}_{\mathrm{spin}}$
\begin{align}
    E'_n = \langle\psi_n| (\hat{H}+\omega\hat{S}_z)|\psi_n\rangle = E_n + \omega \langle \psi_n \vert \hat S_z \vert \psi_n \rangle\;, \label{eq:E-min-rescaled}
\end{align}
and look at $E'_n$ as well as $E_n$ in order to judge how energy is distributed between spin and fermions. 

We then adopt the simple heuristic procedure of setting $N=L/2$ for half-filling, and selecting the $|\psi_n\rangle$ with the lowest $E'_n$ (which is generally not the state of lowest $E_n$). This state should be one which is as close as possible to having all fermions in the lowest energy band. There may well be other states as well with that property, differing in the precise state of the spin, but since the RMM ground state energy is lower for larger $J g$ and $\Delta$, we can expect our lowest $E'_n$ state to have the largest spin projection in the $(x,y)$ plain, and thus the smallest $\langle\hat{S}_z\rangle$. 
So we consider the minimum $E'_n$ state to be the most promising state for our system to behave as an autonomous Thouless pump state, in spite of quantum fluctuations and back-action, and we therefore focus on it. As will be shown later on, this state possesses a finite particle-hole gap for fermion excitations and thus shows the "gapfulness of the fermions" we required above for a topological pump to work.

\subsection{Fermion transport}

Our Hamiltonian $\hat{H}$ implies that the expected particle current from site $l$ to site $l+1$ in the lattice (with $l=L+1$ identified with $l=1$ in the ring) is slightly different from the RMM case (\ref{RMMJ}):
\begin{align}
    \mathcal{J}_{l} =  -iJ\Bigl\langle (1 + (-1)^{l}g\hat S_x)\bigl( \hat{c}_{l+1}^{\dagger}\hat{c}_{l} - \hat{c}_{l}^{\dagger}\hat{c}_{l+1}\bigr) \Bigr\rangle\, .
\end{align}
Since we are considering an eigenstate of $\hat{H}$, this current is time-independent---the externally imposed cycle period $T$ of the Rice-Mele Thouless pump does not apply to our autonomous system. In the semi-classical limit, however, a quantum energy eigenstate is essentially a superposition of all points in a classical trajectory, and so our time-independent state can be pictured as containing all phases of a classical pumping cycle at once, in quantum superposition. In this sense, the time evolution of the spin is still a meaningful aspect of the time-independent quantum state, and we can suppose the spin to be precessing at a rate which must be close to $\omega$, at least for $S$ and $\omega$ large enough that the back-action of the fermions on the spin precession is small. This suggests that the pumping period $T$ which is achieved by the dynamical spin driving should be close to $2\pi/\omega$.

\begin{figure}[H]
    \centering
    \includegraphics[width=0.8\linewidth]{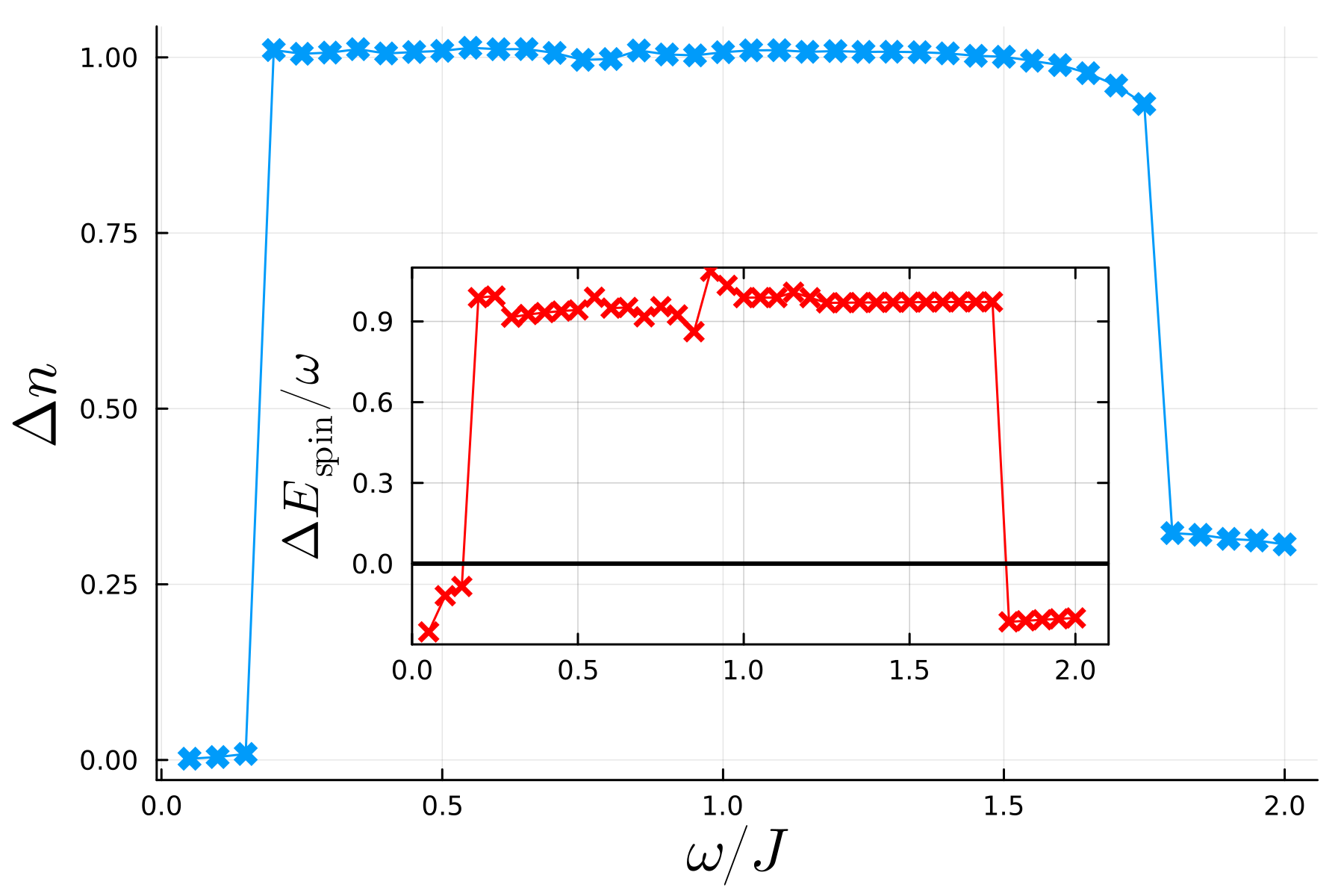}
    \caption{Fermion transport (blue crosses) for a period $T=2\pi/\tilde \omega$, where $\tilde \omega = \omega +\delta \omega$ is a (slightly) renormalized oscillation frequency, explained in Sect. \ref{sect:spin-dynamics}, for $S=10$, $L=8$ and $\eta = 0.95$ obtained by exact diagonalization. One observes a sharp jump at some minimum value $\omega_\textrm{min}$ from a non-transporting region to a phase with quantized transport. At some large value of $\omega/J$ the transport breaks down again. At the two transition points the spin-excitation gap \eqref{eq:spin-gap}, shown in red, closes.
    }
    \label{fig:transport}
\end{figure}

For fermion number $N=L/2$ (half-filling) and in the state $\vert \psi_n\rangle$ with minimal $E_n^\prime$, then, we have numerically calculated (with exact diagonalization) the transport $\Delta n = \mathcal{J}_l T$, for $T\approx 2\pi/\omega$. Within a range of magnetic field strengths $\omega$ this $\Delta n$ remains very close to one; outside this range, $\Delta n$ falls abruptly: see Fig.~\ref{fig:transport}, where we have plotted $\Delta n$ versus $\omega/J$ for lattice size of $L=8$ and spin $S=10$.
More precisely, we use
$T=2\pi/\tilde \omega$, where $\tilde \omega = \omega +\delta \omega$ and $\delta \omega$ is a small correction, whose origin will be explained in the following section. 

The sharp jumps in the transported number of particles are reminiscent of a 
quantum phase transition
in the chosen excited state. While particle-hole excitations of the fermions are always gapped in the chosen eigenstate, spin excitations become gapless at the transitions points, i.e., there is a sudden sign change of the spin excitation energy gap
\begin{equation}
    \Delta E_\textrm{spin}(\omega) = \langle \hat S^+ \psi_n \vert H \vert \hat S^+ \psi_n\rangle - \langle \psi_n\vert H\vert \psi_n\rangle. 
    \label{eq:spin-gap}
\end{equation}
%
$\Delta E_\textrm{spin}$ is also shown in the inset of Fig.~\ref{fig:transport}.

\subsection{Transport phase diagram}

\begin{figure}[H]
    \centering
    \includegraphics[width=\linewidth]{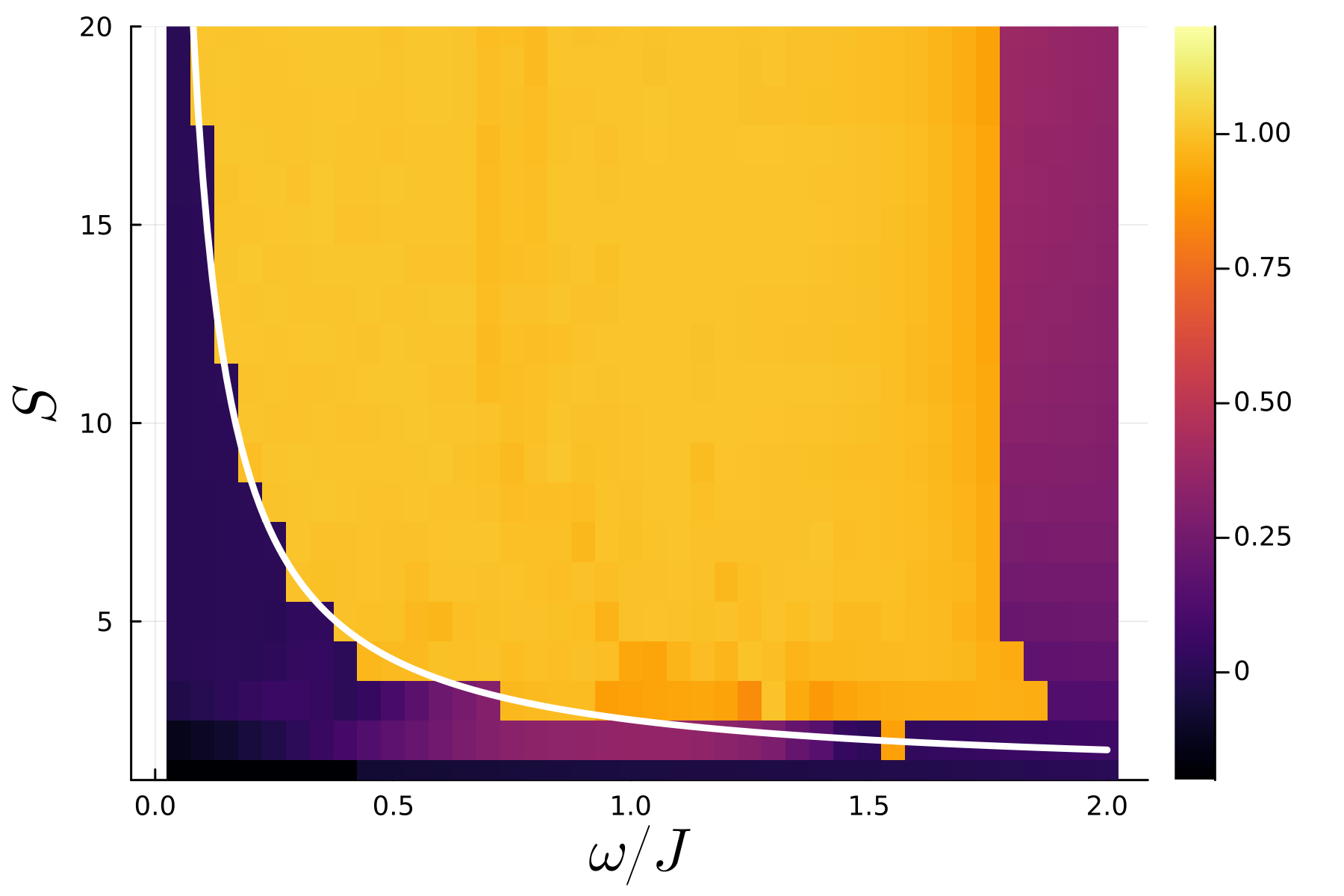}
    \caption{Accumulated transport $J$ at site $l$ for corresponding periods $T=2\pi/\tilde \omega$ for $L=8$ lattice sites and $\eta = 0.95$. For weak magnetic fields $\omega/J$ and/or small spin values the backaction from the fermions stops the transport. The critical spin-value can be described with a $\omega_\textrm{crit}^{(0)}/J \propto S^{-1} $ dependency. The sudden drop for strong magnetic fields corresponds to the loss of adiabaticityin a semiclassical description. 
    }
    \label{fig:phaseDiagram}
\end{figure}

To see the role played by the spin size $S$ in these apparent transitions, in Figure~\ref{fig:phaseDiagram} we have plotted $\Delta n$, for fixed lattice size $L=8$, as a function of $S$ and $\omega/J$, again by exact diagonalization. 
We may recognize:
\begin{itemize}
    \item[(i)] There is sharp transition from a region of exactly vanishing transport to a region of almost perfectly quantized transport with $\Delta n=1$ in a period $T=2\pi/\tilde\omega$, when the magnetic field
    exceeds a certain minimum value $\omega_\textrm{min}\propto 1/S$.
\item[(ii)] Above an upper critical value of the magnetic field $\omega/J \gtrapprox 1.75$ the quantized transport breaks down, which becomes independent on the value of $S$ for large $S$.
\end{itemize}
It is remarkable that there is an extended parameter regime in which the system shows integer-quantized transport in an \textit{excited eigenstate} of the coupled spin-fermion Hamiltonian, with a current fixed by the external magnetic field in $z$ direction. This constitutes an \textit{autonomous pump with quantized  particle transport}. We will argue in the following Section that this quantized transport has topological origin and  is robust against perturbations, and we will give an intuitive explanation for the quantum phase transition at 
$\omega = \omega_\textrm{crit}^{(0)}$.

\subsection{Spin properties} \label{sect:spin-dynamics}

As we have seen in Sect.~\ref{sect:mean-field}, in the semiclassical limit the central spin performs a rotation about the $z$-axis, which corresponds to a cyclic change of parameters of an effective Rice-Mele model causing a topological transport of fermions. 
%
\begin{figure*}
       \includegraphics[width=0.95\textwidth]{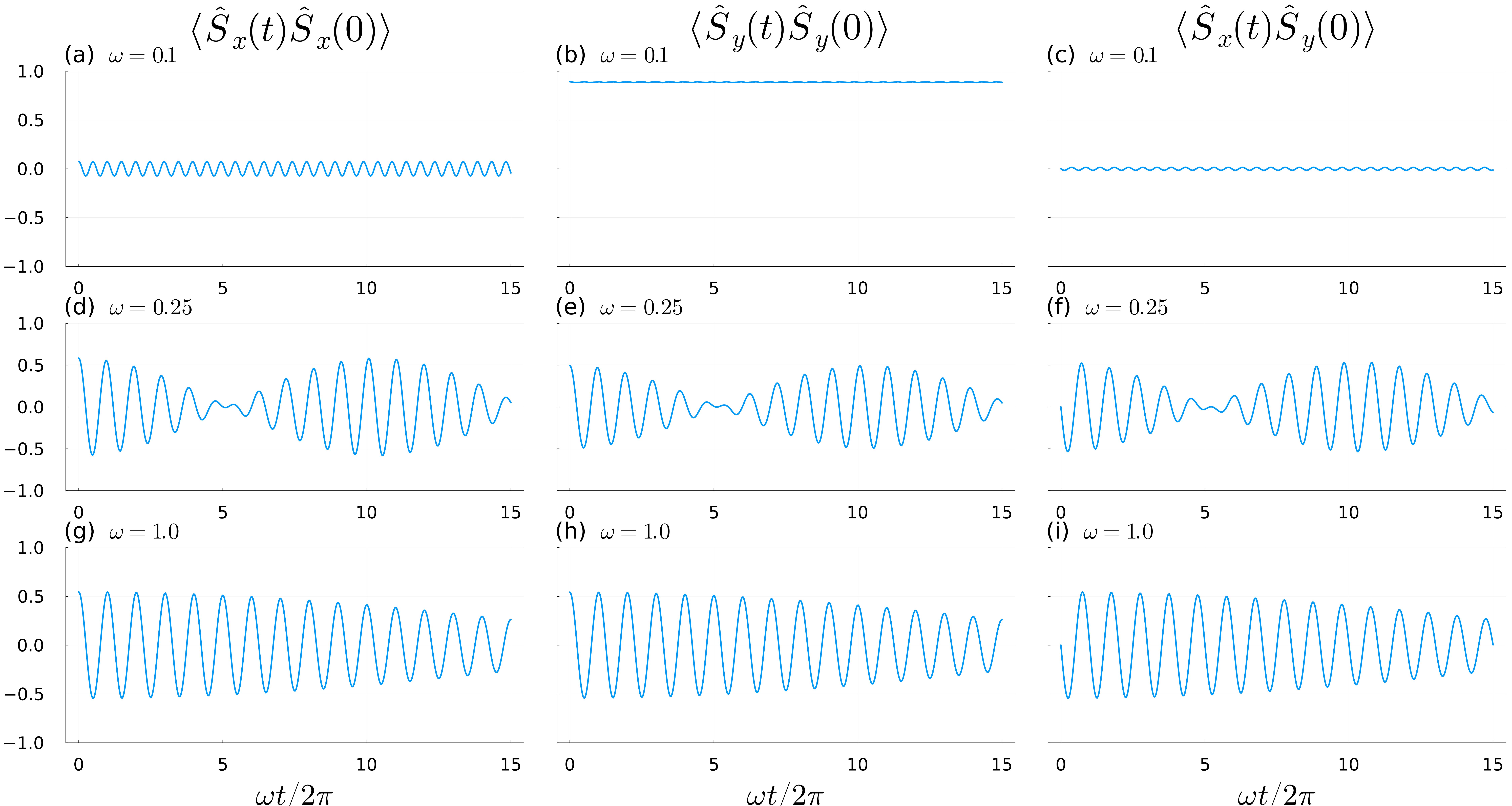}
    \caption{Spin correlations $\langle S_x(t) S_x(0) \rangle$ (left), $\langle S_y(t) S_y(0) \rangle$ (middle) and $\langle S_x(t) S_y(0) \rangle$ (right) in units of $S^2$ for $S=10$, $L=8$ sites, $\eta = 0.95$ and different values of $\omega / J$ . In the non-transporting phase shown in (a), (b) and (c), the spin precession does not encircle the $z$-axis. In the transporting phase 
    just above the phase transition at $\omega_\textrm{min}$, shown in (d), (e) and (f), the spin rotates around the $z$-axis with 
    fast precession due to strong backaction of the fermions. This backaction becomes smaller for larger values of $\omega$, as shown in (g), (h) and (i).
    }
    \label{fig:timeCorrelationSpin}
\end{figure*}
%
In order to see if this holds true in the full model, which includes quantum fluctuations and the backaction of the fermions onto the spin, we must examine the spin dynamics in the eigenstate with lowest expected fermion energy $E^\prime_n$ as defined in Eq.~\eqref{eq:E-min-rescaled}.
Since the Hamiltonian is time-independent, the expectation values of $\hat S_x$, $\hat S_y$ and $\hat S_z$ are constant in time and we find
\begin{align}
    \langle \hat S_x \rangle = &\langle \hat S_y \rangle = 0 \hspace{.5cm} \mathrm{and}  \hspace{.5cm} \langle \hat S_z \rangle \approx 0.
\end{align}
The $x$- and $y$-component can be shown to always be zero in this system for symmetry reasons: For periodic boundary conditions and even number of sites $L$ the Hamiltonian \eqref{eq:mbhamiltonian} is invariant under simultaneous translation by one lattice site and $\hat S_x \to -\hat S_x$, $\hat S_y\to -\hat S_y$, i.e. a rotation of the spin about the $z$ axis by $\pi$, which implies vanishing expectation values $\langle \hat S_{x,y} \rangle=0$.
The $z$-component is not necessarily zero, but numerical calculations show that it
is approximately zero in our chosen state.
The variance of $\Delta S_x$ and $\Delta S_y$ is close to the spin value $S$. 

In order to get some information about the spin-\textit{dynamics}, therefore, we must consider two-time correlations.
Specifically, we calculate the two-time correlations of the $S_x$ and $S_y$ components: $\langle \hat S_\mu(t) \hat S_\mu(0) \rangle$, and the $x-y$ cross-correlation $\langle \hat S_x(t)\hat S_y(0)\rangle$ where it holds $\hat S_\mu(t) = \exp(-i H t) \hat S_\mu(0) \exp(i H t)$. 
These correlation functions are shown in Fig.~\ref{fig:timeCorrelationSpin} for three different values of the magnetic field $\omega$, as obtained by exact diagonalization with spin $S=10$ and lattice size $L=8$.
For large values of $\omega/J$, see Figs.~(g) - (i), both two-time correlations oscillate around zero
and the cross-correlation oscillates with a phase shift of $\pi/2$. This corresponds to a rotation of the spin about the $z$-axis with frequency $\tilde\omega\approx \omega$. 
When $\omega$ decreases one observes a beat note and a small modification $\delta \omega$ of the main oscillation frequency, see Figs.~(d) - (f). This can be understood as a precession of the rotation axis of the spin due to the back-action of the fermions, which cause an additional (small) effective magnetic field in the $(x,y)$-plane.
Below a minimum value $\omega_\textrm{min}$ the total effective magnetic field in the $(x,y)$-direction is larger than the total field's $z$-component, and so the spin no longer precesses about the $z$-axis; see Figs.~(a)-(c).

\section{Topological origin of quantized transport}

We now argue that the quantized transport that we have observed in numerical simulations is of topological origin. We first give an explanation for the transition from the non-transporting to a transporting phase at $\omega=\omega_\textrm{crit}^{(0)}$ and $\omega=\omega_\textrm{crit}^{(1)}$. For this, we have to restrict ourselves to the limit of small quantum fluctuations. Subsequently, we will show for the limit of large spin $S$ that the transport quantization in the transport phase can be traced back to a topological invariant. 

\subsection{Mean-field decoupling approximation}

In order to understand the coupled spin-fermion dynamics, we will consider the limit of small
quantum fluctuations, where the full Hamiltonian, Eq.~\eqref{eq:mbhamiltonian}, is treated by
a mean-field decoupling of spin and fermion degrees of freedom. To this end, we restrict ourselves to many-body states for which second order fluctuation terms in spin ($\hat S$) and fermion ($\hat F$) operators can be ignored in the Hamiltonian, \textit{i.e.} 
\begin{equation}
   \delta \hat S \, \delta \hat F =  \bigl( \hat S - \langle \hat S\rangle\bigr) \bigl( \hat F - \langle \hat F\rangle\bigr) \approx 0.
\end{equation}
In this approximation, the fermion and spin dynamics are both governed by effective time-dependent Hamiltonians
\begin{eqnarray}
    H^\textrm{MF}_\textrm{f}(t) &=& - \frac{J}{2} \sum_{j=1}^L \Bigl( 1 + (-1)^j g \langle \hat{S}_x(t)\rangle\Bigr) (\hat{c}_j^\dagger \hat{c}_{j+1} + h.c.)\nonumber \\
    && + J \sum_{j=1}^L (-1)^j g \langle\hat{S}_y(t)\rangle \hat{c}^\dagger_j \hat c_j, 
   \label{eq:eff-fermion} \\
     H^\textrm{MF}_\textrm{spin}(t) &=& - \omega \hat S_z - B_x(t) \hat{S}_x \label{eq:eff-spin}- B_y(t) \hat{S}_y.
    \label{eq:effective_magnetic_field}
\end{eqnarray}
Here the effective magnetic field in $x$ and $y$ direction is given by current imbalance $\hat I$ and the unit cell
polarization $\hat P$
\begin{eqnarray}
     B_x(t) &=& g J \langle \hat I\rangle =  
     \frac{ gJ}{2}\sum_j (-1)^j \bigl(\langle \hat c_j^\dagger\hat c_{j+1}\rangle + c.c.\bigr), \nonumber\\
    B_y(t) &=& g J \langle \hat P\rangle =  g J \sum_j(-1)^{j+1} \langle \hat c_j^\dagger \hat c_j\rangle.
\end{eqnarray}
The effective fermion Hamiltonian, Eq.~\eqref{eq:eff-fermion}, is formally identical to the Rice-Mele model and 
shows a quantized topological transport if and only if the projection of the spin onto the $(x,y)$-plane encircles the origin. Since the spin projections are self-consistently determined by the Heisenberg equations 
\begin{equation}
    \frac{d \langle\hat{\vec S}\rangle}{dt} = - \vec B(t) \times \langle\hat{\vec S}\rangle,
    \quad \vec{B}(t) = \left(
    \begin{array}{c}
          g J \langle \hat I(t)\rangle \\
          g J \langle \hat P(t)\rangle\\
          \omega
    \end{array}
    \right)\;,
\end{equation}
this is the case if the $z$-component of the magnetic field $\omega$ is the largest component. When we include the back-action of the fermions onto the central spin as part of the total effective magnetic field, however, the in-plane components of this total field can become large enough that the spin precession may no longer encircle the $z$-axis. To prevent that from happening, the magnetic field $z$-component given by $\omega$ must be strong enough to dominate the fermionic back-action. Since $\vert \langle \hat I\rangle \vert \le N $ and $\vert \langle \hat P\rangle\vert \le N$, we can estimate this threshold $\omega$ to be
\begin{equation}
   \frac{\omega_\textrm{crit}^{(0)}}{J} > g N = \eta \frac{N}{S}
\end{equation}
where in the last expression we have used $g= \eta S^{-1}$. This accounts for the transition to a non-pumping state at the lower $\omega/J$ as shown in Figs.~\ref{fig:transport} and \ref{fig:phaseDiagram}.

On the other hand, if $\omega$ is larger than the energy gap of the effective Rice-Mele model, Eq.~\eqref{eq:eff-fermion}, topological transport is expected to break down, as the time modulation is no longer adiabatic. This criterion gives the upper limit
\begin{align}
    \frac{\omega_\mathrm{crit}^{(1)}}{J} = \frac{\Delta E_\textrm{gap}^\textrm{MF}}{J} = 2\eta  \sqrt{\frac{\langle \hat S_x \rangle^2}{S^2} + \frac{\langle \hat S_y \rangle^2}{S^2}}\le 2 \eta\;,  
\end{align}
which agrees very well with the numerically obtained upper critical value of $\omega/J$ at which $\Delta n$ jumps back from unity to a small, non-quantized value in Fig.\ref{fig:transport}.

\subsection{Limit of large $S/N$: topological invariant and quantized transport}

Let us now consider the limit in which the spin is not only large, but also much larger than the total number of fermions, $S\gg N$. We also assume that $\omega/J$ is not too small, $\omega/J \gg N/S$, so that we are deep within the transporting regime, where the back-action of the fermions onto the spin dynamics must be small.

We first apply a Holstein-Primakoff mapping from spin to bosons $\hat a$, and $\hat a^\dagger$, with the vacuum state corresponding to the eigenstate of $\hat S_z$ with lowest eigenvalue $-S$:
\begin{eqnarray*}
    &\hat S_- =\Bigl( \sqrt{2 S - \hat a^\dagger \hat a}\Bigr)\,   \hat a,\qquad \hat S_+ = \hat a^\dagger\, \Bigl(\sqrt{2 S - \hat a^\dagger \hat a}\Bigr) &,\\ 
    &\hat S_z = \hat a^\dagger \hat a - S. &
\end{eqnarray*}
Moreover we use a number-phase decomposition of the oscillator ladder operators
\begin{equation}
    \hat a = \sqrt{\hat n}\, e^{i\hat \phi},\qquad \hat a^\dagger = e^{-i\hat \phi} \sqrt{\hat n},
\end{equation}
where $e^{i\hat \phi} = \sum_{n=0}^\infty \vert n\rangle \langle n+1\vert$. We note that $\hat \phi$ is in general not a Hermitian operator since $[e^{i\hat \phi},e^{-i\hat \phi}] = \vert 0\rangle\langle 0\vert$ and not zero. However, when only states are relevant with high occupation numbers, $\hat \phi$ can be used as approximate Hermitian phase operator, conjugate to the number operator $\hat n$:
\begin{eqnarray}
    \bigl[e^{i\hat \phi}, \hat n\bigr] = e^{i\hat \phi},\qquad \textrm{or}\qquad \bigl[\hat \phi,\hat n\bigr] = i.
\end{eqnarray}
which holds in the subspace of states with large $n$.
With this the Hamiltonian Eq.~\eqref{eq:mbhamiltonian} can, apart from an irrelevant constant term, be written as 
\begin{align}
    & H =  -\omega \hat n \label{eq:mbhamiltonian-2}\\
   & - \frac{J}{2} \sum_{j=1}^L \Bigl( 1 + (-1)^j \eta
    \sqrt{\Bigl(2-\frac{\hat n}{S}\Bigr)\frac{\hat n}{S}} \cos{(\hat \phi)}
    \Bigr) (\hat{c}_j^\dagger \hat{c}_{j+1} + h.c.) \nonumber\\
    &\quad + J \sum_{j=1}^L (-1)^j \eta \sqrt{\Bigl(2-\frac{\hat n}{S}\Bigr)\frac{\hat n}{S}}\sin(\hat \phi) \, \hat c_j^\dagger \hat c_j. \nonumber
\end{align}
Since in the transport regime the $z$ component of the spin is close to zero, \textit{i.e.} $\hat n \approx S$, we furthermore replace the square root terms with unity. With this approximation we neglect the back-action of the fermion system onto the spin dynamics, which should be accurate for $\omega/J \gg \omega_\textrm{min}/J= \eta \frac{N}{S}$. 
This then yields the approximate Hamiltonian
\begin{align}
    H \approx  -\omega \hat n &- \frac{J}{2} \sum_{j=1}^L \Bigl( 1 + (-1)^j \eta \cos{\hat \phi}
    \Bigr) (\hat{c}_j^\dagger \hat{c}_{j+1} + h.c.) \nonumber\\
    &+ J \sum_{j=1}^L (-1)^j \eta \sin\hat \phi \, \hat c_j^\dagger \hat c_j, \label{eq:mbhamiltonian-3}
\end{align}
In this limit the angle $\hat \phi$ has a simple dynamics $d \hat \phi / dt = \omega$. 

The system has discrete translational invariance with a unit cell of two lattice sites. Thus 
the lattice momentum $k= \pi n/L$, ($n=(-L/2,\dots,L/2)$) is a conserved quantity, and after 
a discrete Fourier transformation the Hamiltonian can be written as
\begin{eqnarray}
   && H = - \omega \hat n + \sum_k \left(\begin{array}{c}
         \hat c_A(k) \\
         \hat c_B(k)
    \end{array}\right)^\dagger
    h(k,\hat \phi)
     \left(\begin{array}{c}
         \hat c_A(k) \\
         \hat c_B(k)
    \end{array}\right)
   ,\nonumber
\end{eqnarray}
where the index $A$ denotes the sublattice with even $j$ and $B$ the sublattice with odd $j$.
$h(k,\hat\phi)$ is a $2\times 2$ matrix in fermion space and depends on lattice momentum $k$
and the angle variable $\hat \phi$.

\begin{eqnarray}
    && h (k,\hat \phi) =    J\eta \sin\hat \phi\,  \sigma_z + \frac{J}{2}\Bigl(1 -\eta \cos \hat \phi\Bigr) \sin k \, \sigma_y\nonumber\\
    && \quad- \frac{J}{2}\Bigl( 1+\eta \cos\hat \phi +(1-\eta \cos\hat\phi) \cos k \Bigr)\, \sigma_x
\end{eqnarray}
For any given eigenvalue $\phi$ of $\hat \phi$, $h$ has two eigenstates corresponding to two bands which we denote as "-" and  "+", and with energies ${\cal E}_\pm(k,\phi) = \pm \,  {\cal E}(k,\phi)$ 
\begin{eqnarray}
    {\cal E}(k,\phi) &=& J\biggl[\eta^2 \sin^2\phi + \frac{1}{2}(1+\eta^2\cos^2\phi) + \\
    && \qquad \frac{1}{2}(1-\eta^2\cos^2\phi) \cos k
    \biggr]^{1/2}.
\end{eqnarray}

For $\eta \ne 0$ there is a finite band gap.
Thus for $\omega \to 0$ the ground state of the system at half filling is a tensor product of an eigenstate $\vert \phi_0\rangle$ of the angle operator and a band insulator of fermions where each $k$ state in the lower ("-") band is occupied by one fermion $\prod_k\vert \psi_-(k,\phi_0)\rangle = \prod_k \hat c_-^\dagger(k,\phi_0) \vert 0\rangle$. $\phi_0$ is the eigenvalue for which the
total energy of the lower band is minimal. 

In order to calculate the current for finite but small values of $0<\omega/J \ll 1$, we apply a time-independent perturbation theory in $\epsilon=\omega/J$. We cannot take the eigenstate $\vert \phi_0\rangle$ in tensor product with $\vert \psi_-(k,\phi_0)\rangle$ as zeroth-order energy eigenstate, however, since $\omega \hat n$ is not a small perturbation to a $\hat \phi$ eigenstate.
Instead,  we have to use an ansatz where all values $\phi$ occur with equal weight
\begin{eqnarray}
    &&\vert \Psi\rangle =\int_0^{2\pi}\!\!\! d\phi\, e^{-i \gamma(\phi)}\biggl[\prod_k \vert \psi_-(k,\phi)\rangle + \label{eq:psi-perturbation}\\
    &&+ \epsilon \sum_k C_k(\phi) \vert \psi_+(k,\phi)\rangle \prod_{k^\prime \ne k} \vert \psi_-(k^\prime,\phi)\rangle + {\cal O}(\epsilon^2)\biggr]\vert \phi\rangle,\nonumber
\end{eqnarray}
where $\gamma(\phi)$ is a phase factor, for which we set 
\begin{equation*}
    \gamma(\phi) = \frac{1}{\epsilon}\int_0^\phi \!\! d\phi^\prime \, K(\phi^\prime)
\end{equation*}
To solve $H\vert \Psi\rangle = E\vert \Psi\rangle$ in perturbation theory, we expand the energy and phase in powers of $\epsilon$:
\begin{equation*}
    E = \sum_{n=0}^\infty \epsilon^n E_n,\qquad K(\phi) =  \sum_{n=0}^\infty \epsilon^n K_n(\phi).
\end{equation*}
Substituting this into the stationary Schr\"odinger equation eventually yields
\begin{equation}
    C_k(\phi) = \frac{i \langle \psi_+(k,\phi)\vert \partial_\phi \vert \psi_-(k,\phi)\rangle }{ {\cal E}_+(k,\phi)- {\cal E}_-(k,\phi)}.
\end{equation}
The total current of the fermion system is given by
\begin{eqnarray}
    {\cal J} = \frac{1}{2\pi} \int_\textrm{BZ}\! \! dk \,  \Bigl\langle \Psi\Bigr\vert \frac{\partial h(k,\hat \phi)}{\partial k} \Bigl\vert \Psi\Bigr\rangle.
\end{eqnarray}
Inserting the first order correction to the eigenstate, Eq.~\eqref{eq:psi-perturbation}, and 
making use of $\langle \psi_-(k)\vert \partial_k h\vert \psi_+(k)\rangle =\langle \partial_k \psi_-(k)\vert \psi_+(k)\rangle ({\cal E}_+(k)- {\cal E}_-(k))$, 
one finds
\begin{equation}
    \frac{{\cal J}}{\omega} = \frac{i}{2\pi}\int_\textrm{BZ}\! \! \! dk\int_0^{2\pi}\!\!\! d\phi \biggl(\Bigl\langle\frac{\partial \psi_-(k,\phi)}{\partial k}\Bigr\vert \frac{\partial \psi_-(k,\phi)}{\partial \phi}\Bigr\rangle - c.c.
    \biggr),
\end{equation}
which is a Chern number and thus the current is integer-quantized in units of $\omega$. This explains the quantized transport of $\Delta n = {\cal J}T = 1$ in a period $T=2\pi/\omega$, observed numerically in the previous section. Since our discussion has presumed the limit in which the back-action of the fermions onto the
dynamics of the spin is negligible, there is no back-action shift of the pumping period and we simply have $\tilde{\omega}=\omega$.

\begin{figure}[H]
    \begin{centering}
        \includegraphics[width=0.9\linewidth]{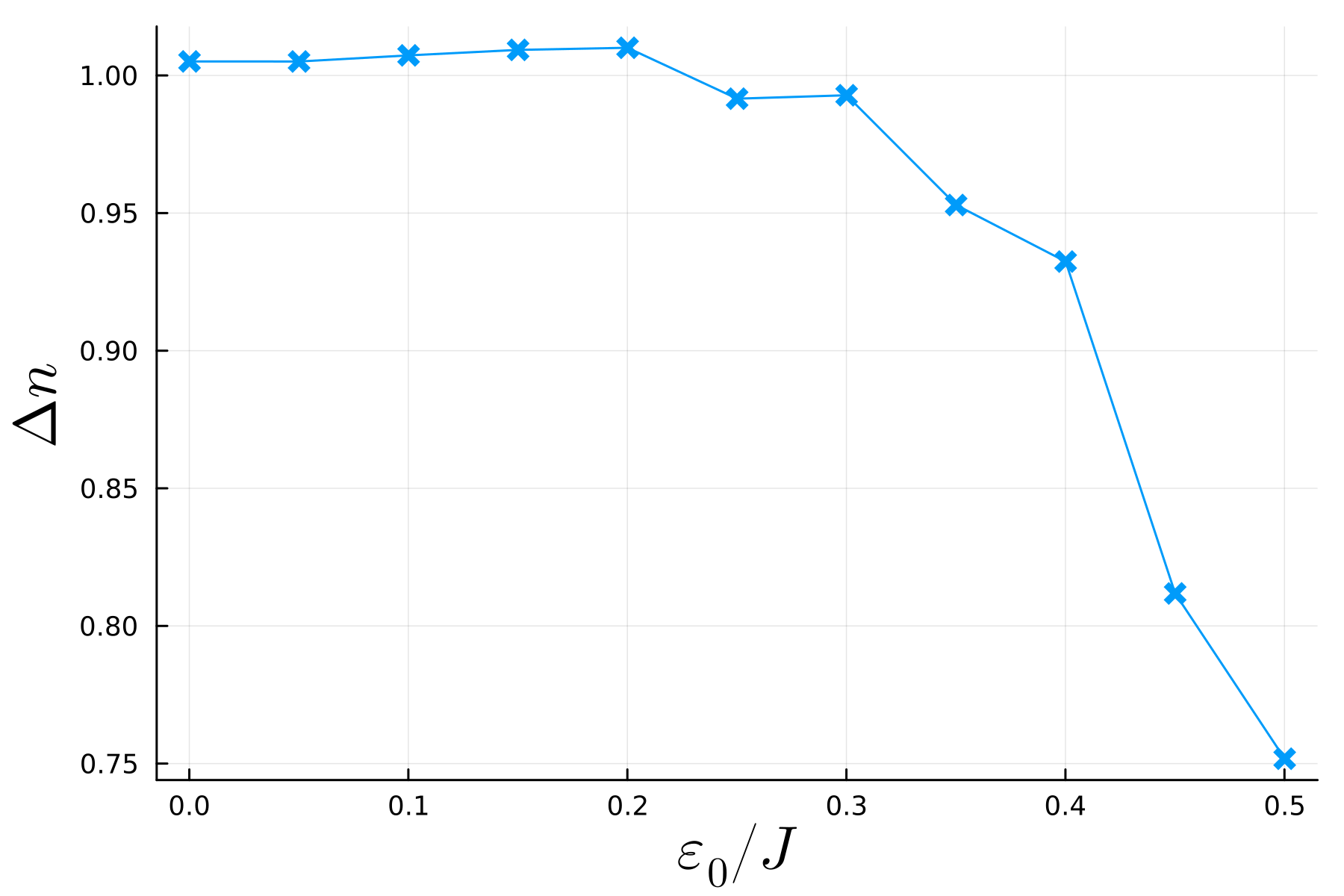}
    \end{centering}
    \caption{Transport over one period of time $T = 2\pi / \tilde \omega$ for $S=10$, $L=8$, $\eta = 0.95$ and $\omega/ J=0.25$ with different amplitudes of disorder $\epsilon_0$, averaged over 50 disorder realizations. One can see the breakdown of transport when the disorder is of the same order as $J$.}
    \label{fig:disorder}
\end{figure}

\section{Robustness of transport}

Thouless pumps are well known to be topologically protected against disorder potentials \cite{Niu_1984}. To show this in our system, we now add time-independent on-site potential terms with random amplitudes  \eqref{eq:mbhamiltonian}:
\begin{align}
    H^\prime &= H + \sum_{j=1}^L \epsilon_j \hat{n}_j.
\end{align}
The $\epsilon_j$ are taken from a box probability distribution with $0 \le \epsilon_j \le \epsilon_0$. 
In Figure \ref{fig:disorder} we show the transport, averaged over 50 disorder realizations, for a given set of parameters $S$, $L$ and $\omega$ while varying the disorder strength $\epsilon_0$. We can see that the transport only breaks down when the disorder strength approaches the same order as the hopping, \textit{i.e.} for  $\epsilon_0 \approx 0.3 J$. This is in agreement with the expected behavior of the original Rice-Mele model. 

An important difference between our system and the Rice-Mele model is the absence of an energy gap of the many-body ground state in the coupled fermion-spin system, see Fig.~\ref{fig:fullspectrum}.
Therefore one might naively conclude that the transport cannot be robust against potential perturbations. 

\begin{figure}[H]
    \begin{centering}
        \includegraphics[width=0.9\linewidth]{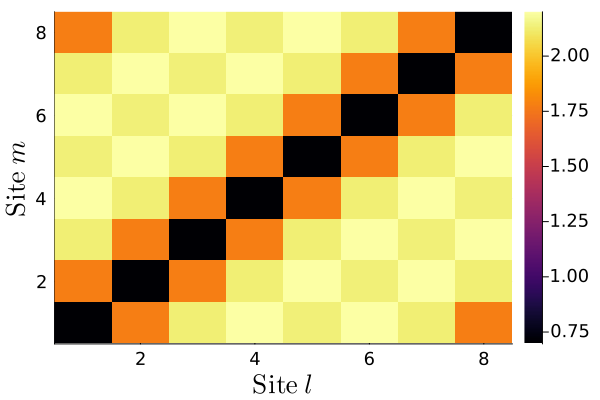}
    \end{centering}
    \caption{Particle-hole-gap $\Delta E_{\mathrm{p-h},l,m}/J$ for a system with $S=10$, $L=8$, $\eta = 0.95$ and $\omega/J = 0.25$. The gap is always large compared to the frequency $\omega/J$, making the state robust against excitations from the Hamiltonian.}
    \label{fig:part-hole-gap}
\end{figure}

This is not the case here, however, as the total energy gap is not relevant here. 
Rather, we have to consider the energy difference between many-body states that are connected by a perturbation acting only on the fermions. To this end, we calculate the particle-hole gap which is the energy that is needed to move one particle from site $l$ to site $m$:
\begin{align}
    \Delta E_{\mathrm{p-h},l,m} = \langle \psi \vert H \vert \psi \rangle - \langle \psi \hat{c}^{\dagger}_{l} \hat{c}_{m} \vert H \vert \hat{c}^{\dagger}_{m} \hat{c}_{l} \psi\rangle.
\end{align}
In Fig.~\ref{fig:part-hole-gap} we have plotted the particle hole gap for $S=10$, $L=8$ and $\omega / J = 0.25$. One recognizes that the gap is always finite and of the order of $J$. 
%

\section{Summary and outlook}

In the present paper we have shown that a robust quantized particle particle current can be induced in a time-independent fermionic system 
coupled to a global spin in a constant magnetic field.
This constitutes a minimal model of an autonomous topological pump, which we believe is an important step in constructing self-operating quantum machines by combining topological protection against the influence of perturbations and noise with autonomous operation, \textit{i.e.} without need of external control. 
Analyzing the spectrum of the time-independent Hamiltonian, we found that there are excited eigenstates, corresponding to insulating states of the fermions, which display quantized transport on a timescale governed by the magnetic field. Although the total Hamiltonian is gapless, particle-hole excitations of the fermions are gapped.
The quantized transport can be understood intuitively, by looking at the decoupled system, where the spin displays Larmor precession enforced by the magnetic field. We extend this understanding with a self-consistent mean-field theory of decoupled systems.
Upon increasing the strength of the magnetic field,
a sharp transition from a non-transporting 
phase to one with integer-quantized transport occurs.
Increasing the strength even further, a second transition into a phase with 
non-quantized transport eventually takes place. 
While the chosen eigenstate displays a finite particle-hole gap for all values of the magnetic field, the two transition points coincide with sudden zero-crossings of the spin excitation gap. 
In the first non-transporting phase the backaction of the fermions onto the dynamics of the spin is so strong that it suppresses the transport. 
We show the topological origin of the quantized transport in the intermediate phase by considering the limit of very large spin-values $S$.
Specifically, we show the existence of a topological invariant, equivalent to a Chern-number, with which we can explain the quantization of particle transport.
To verify the topological protection of the transport in this parameter regime, we furthermore showed its robustness against small potential disorder.
The breakdown of quantized transport for large values of the magnetic field can be understood intuitively
by considering the mean-field limit, where the fermion Hamiltonian becomes time-dependent due to the 
rotation of the spin with an angular frequency given by the magnetic field strength. If this frequency becomes too large,
adiabaticity breaks down and the transport is no longer quantized. In the full model, the selection of the specific eigenstate as ''transporting'' eigenstate is then no longer sensible in this parameter regime. 

We note that our autonomous topological pump still holsa some conceptual challenges. For example, in the conventional Thouless pumps
perfect topological transport occurs for a few special states of the many-body fermion systems, namely states with fully
occupied bands. Our coupled system can be expected to show quantized transport in different eigenstates. Many
different states 
of the spin can all drive the pump, because it is topological, and so an effective band insulator 
of the fermions can now be combined with different possible states of the spin leading to a phase characterized by the same topological invariant. Moreover, we were only able to derive a topological invariant in a limiting case, and the precise nature of the observed transitions is still unclear.
Despite these open questions,
we believe that the combination of quantized topological transport, observed in our minimal model, with an autonomous operation opens a new avenue for the construction of topologically protected quantum machines. 

\ 

\paragraph*{Acknowledgements}
The authors thank Til M\"ohnen for fruitful discussions. Financial support from the DFG through SFB TR 185, Project No. 277625399, is gratefully acknowledged. The authors also thank the Allianz für Hochleistungsrechnen (AHRP) for giving us access to the “Elwetritsch” HPC Cluster. The data in this script has been obtained using the \textit{QuantumOptics.jl} framework \cite{kramer2018quantumoptics}.


\bibliography{bibliography.bib}

\end{document}